\documentclass[prd,nofootinbib,english]{revtex4}

\usepackage{graphicx,float}
\usepackage{amsmath,amssymb,amsfonts}
\usepackage{mathrsfs}
\usepackage{epsfig,color}
\usepackage[thinlines]{easytable}
\usepackage{pdfpages}
\usepackage{array}
\usepackage{cancel}
\usepackage{mathtools}
\usepackage{accents}
\usepackage{subfigure}
\usepackage{enumitem}
\usepackage[dvipsnames]{xcolor}
\usepackage{refstyle}
 \usepackage{hyperref}
\hypersetup{
    colorlinks=true,
    linkcolor=blue,
    filecolor=magenta,
   citecolor=blue
}

\def\be{\begin{equation}}
\def\ee{\end{equation}}
\def\bea{\begin{eqnarray}}
\def\eea{\end{eqnarray}}

\begin{document}
\hfill  USTC-ICTS/PCFT-21-17
\title{Revisiting a parity violating gravity model without ghost instability: Local Lorentz covariance}

\author{Mingzhe Li}
\author{Haomin Rao}
\author{Yeheng Tong}
\affiliation{Interdisciplinary Center for Theoretical Study, University of Science and Technology of China, Hefei, Anhui 230026, China}
\affiliation{Peng Huanwu Center for Fundamental Theory, Hefei, Anhui 230026, China}

\begin{abstract}
Recently, based on the theory of teleparallel gravity, a simple and ghost free parity violating gravity model was proposed in [M. Li, H. Rao, and D. Zhao, J. Cosmol. Astropart. Phys. 11 (2020) 023], where the Weitzenb\"{o}ck connection was adopted for simplifying the calculations but breaks the local Lorentz symmetry explicitly. In this paper, we restore the local Lorentz symmetry of this model by giving up the Weitzenb\"{o}ck condition on the spin connection. With
full local Lorentz covariance, this model is not a pure tetrad theory any more. We also apply the new version of this model to the universe with general Friedmann-Robertson-Walker background. This further generalizes the
studies of [M. Li, H. Rao, and D. Zhao, J. Cosmol. Astropart. Phys. 11 (2020) 023] where only the spatially flat background was considered.
Through the investigations of this paper, we confirm the results obtained in
[M. Li, H. Rao, and D. Zhao, J. Cosmol. Astropart. Phys. 11 (2020) 023] and in addition get some new results.
\end{abstract}

\maketitle

\section{introduction}

Parity violating (PV) gravities attracted a lot of interests in recent years. A famous and frequently studied PV gravity is the so-called Chern-Simons (CS) modified gravity \cite{Jackiw:2003pm,Alexander:2009tp}, in which general relativity (GR) is modified by a gravitational CS term; $S_{CS}\sim \int d^4x \sqrt{-g} \theta(x) \varepsilon^{\mu\nu\rho\sigma}R_{\mu\nu}^{~~~\alpha\beta}R_{\rho\sigma\alpha\beta}$, where $g$ is the determinant of the metric, $\theta(x)$ is a coupling scalar field,  $\varepsilon^{\mu\nu\rho\sigma}$ is the four-dimensional Levi-Civita tensor and $R_{\rho\sigma\alpha\beta}$ is the Riemann tensor constructed from the metric. The CS modified gravity makes a difference between the amplitudes of the left- and right-handed polarized components of gravitational waves (GWs), but there is no difference between their velocities. This is the so-called amplitude birefringence phenomenon. However, the CS gravity suffers from the problem of ghost instability \cite{Dyda:2012rj} which essentially originated from the higher derivatives in the CS term. To circumvent this problem, further extensions to the CS gravity were explored in Ref. \cite{Crisostomi:2017ugk} where several terms including higher derivatives of the coupling scalar field $\theta(x)$ were taken into account. Though each term contains higher derivatives, combining them together in a special way can eliminate the ghost modes. The price one must to pay is that all such kind of models have much more complex forms \cite{Gao:2019liu,Zhao:2019xmm}.

Recently we proposed a new PV gravity model which is healthy and simple in form \cite{PVtele1}.
This model is based on the theory of teleparallel gravity (TG) \cite{Tele}
which is equivalent to GR but formulated in flat spacetime with vanishing curvature and vanishing nonmetricity. The gravity in TG theory is identical to the spacetime torsion. Our model \cite{PVtele1} modified the GR equivalent TG theory by an extra PV term:
$S_{NY}\sim \int d^4x \sqrt{-g} ~\theta(x)\varepsilon^{\mu\nu\rho\sigma}\mathcal{T}_{A\mu\nu}\mathcal{T}^{A}_{~~\rho\sigma}$, here $\mathcal{T}^{A}_{~~\mu\nu}$ is the torsion two form and its exact expression and other details will be presented in the next section. Except for the coupling with a scalar field, this PV term is in fact the reduced Nieh-Yan term \cite{Nieh:1981ww} within the framework of teleparallelism. Different from the CS modified gravity, the PV term in this model hides no higher derivatives and successfully avoids the ghost mode. The Nieh-Yan modification to GR equivalent TG has been also considered in Ref. \cite{Chatzistavrakidis:2020wum} where its consequences in gravitoelectromagnetism was studied with a nondynamical $\theta$ field. Still within the framework of TG theory, generalizations containing more PV couplings beyond Nieh-Yan were made in Ref. \cite{Hohmann:2020dgy} where the cosmological dynamics in the Friedmann-Robertson-Walker (FRW) universe was derived.

In Ref. \cite{PVtele1}, our PV gravity model has also been applied to cosmology, where two additional conditions were set by hand. The first one is the so-called Weitzenb\"{o}ck condition, i.e., the spin connection vanishes, $\omega^A_{~B\nu}=0$. This had been frequently adopted for simplifying the calculations in many studies of TG theory and its extensions, such as $f(\mathbb{T})$ models \cite{fT}. The unsatisfactory aspect of taking the Weitzenb\"{o}ck connection is the explicit violation of the local Lorentz symmetry. This is not a difficulty for TG theory because fixing any possible connection there (including the Weitzenb\"{o}ck one) which satisfies the teleparallel constraints will only contribute a surface term in the action; this will not affect the equation of motion. It is not so, however, for modified TG theories in which the equation of motion indeed depends on the spin connection. The second additional condition adopted in Ref. \cite{PVtele1} is that the application to cosmology was limited to the case where the universe has a spatially flat FRW background. With these, we have studied the background evolution and computed the linear cosmological perturbations. We found that the PV modification has no effect on the background evolution but indeed affects the perturbations. For tensor perturbations the left- and right-handed polarized components of GWs have different propagating velocities, but their amplitudes are the same. This phenomenon is called velocity birefringence. For scalar perturbations,  the coupled scalar field $\theta(x)$ surprisingly does not contribute an independent dynamical component at the linear level---it behaves like an imperfect fluid and causes viscosities in the stress-energy tensor.

In this paper, we will revisit our PV gravity model \cite{PVtele1} and its applications in cosmology by relaxing the two additional conditions mentioned above.
Without the Weitzenb\"{o}ck condition, we will leave the spin connection arbitrary as long as the teleparallel constraints (no curvature and no nonmetricity) are satisfied. This restores the local Lorentz symmetry of our model and in this new version both the tetrad and the spin connection are treated as fundamental variables in the action principle. Secondly, we will apply the new version of our model to the universe which has a general FRW background, not just the spatially flat one. We will focus on the linear cosmological perturbations. With these improvements and generalizations, we will confirm the results obtained in Ref. \cite{PVtele1} and get some new results: for instance, the linear perturbation of the scalar field $\theta(x)$ is indeed counted as a dynamical degree of freedom in the case of a curved FRW background.

This paper is organized as follow. In Sec. \ref{TG and PV extension},
after a brief review of the PV gravity model proposed in \cite{PVtele1}, we consider its new version in which the local Lorentz symmetry is restored by giving up any extra constraint on the connection. Then we apply this new version of our model with full local Lorentz covariance to cosmology in the next sections.
In Sec. \ref{cosmology background}, we study the evolution of the background which is a general FRW spacetime. We investigate case by case the effects of the PV term on the evolution equations for the flat, closed and open universes. In section \ref{perturbation}, we discuss the linear perturbations around the FRW background and the gauge transformations.
This ensures that some gauge conditions are available to simplify subsequent calculations.
In Sec. \ref{quadratic action}, we calculate the quadratic actions for scalar and tensor perturbations. We will confirm the results obtained in \cite{PVtele1} and show some new results---among them the most important one is that
the dynamical scalar field $\theta$ acquires its independent dynamics at the linear-perturbation level around the spatially curved FRW background.

\section{the healthy party violating gravity model based on TG theory}\label{TG and PV extension}

In this paper, we adopt the unit $8\pi G=1/M_p^2=1$ and the convention of most negative signatures for metrics. The local space tensor indices  are denoted by $A, B, C,...=0, 1, 2, 3$ and by $a, b, c,...=1, 2, 3$ when limiting to spatial components. They are are lowered and raised by the Minkowski metric $\eta_{AB}$ and its inverse $\eta^{AB}$. The spacetime tensor indices are denoted by Greek $\mu, \nu, \rho,...=0, 1, 2, 3$ and by Latin $i, j, k,...=1, 2, 3$ when limiting to spatial components. These are lowered and raised by the spacetime metric $g_{\mu\nu}$ and its inverse $g^{\mu\nu}$. The tetrad is denoted as $e^A_{~\mu}$ and considered as the square root of the metric, i.e., $g_{\mu\nu}=\eta_{AB}e^A_{~\mu}e^B_{~\nu}$.
The equation $\hat{\Gamma}^{\mu}_{~\rho\sigma}=e_A^{~\mu}(\partial_{\rho}e^A_{~\sigma}+\omega^A_{~B\rho}e^B_{~\sigma})$ relates the affine connection $\hat{\Gamma}^{\mu}_{~\rho\sigma}$ and the spin connection $\omega^{A}_{~B\mu}$. Furthermore, we have the Levi-Civita connection from the metric, $\Gamma^{\mu}_{~\rho\sigma}=(1/2)g^{\mu\nu}(\partial_{\rho}g_{\sigma\nu}+\partial_{\sigma}g_{\rho\nu}-\partial_{\nu}g_{\rho\sigma})$ and its associated covariant derivative operator $\nabla=\partial+\Gamma$. The Levi-Civita tensor $\varepsilon^{\mu\nu\rho\sigma}=(1/\sqrt{-g})\epsilon^{\mu\nu\rho\sigma}$ is related with a totally antisymmetric symbol $\epsilon^{\mu\nu\rho\sigma}$ which satisfies $\epsilon^{0123}=1$.  This will be used to define the dual tensors.

 GR equivalent TG theory is formulated in a flat spacetime with metricity, so both the curvature two form and the nonmetricity tensor vanish, i.e.,
$\hat{R}^A_{~B\mu\nu}=\partial_{\mu}\omega^A_{~B\nu}-\partial_{\nu}\omega^A_{~B\mu}+\omega^A_{~C\mu}\omega^C_{~B\nu}-\omega^A_{~C\nu}\omega^C_{~B\mu}=0$ and $Q_{\rho}^{~\mu\nu}=\hat{\nabla}_{\rho}g^{\mu\nu}=0$. The metricity requires that $\omega_{AB\mu}=-\omega_{BA\mu}$; it together with the vanishing curvature dictate that the spin connection can be generally expressed as follows:
\be\label{omega}
\omega_{~B \mu}^{A}=\left(\Lambda^{-1}\right)_{~C}^{A} \partial_{\mu} \Lambda_{~B}^{C}~,
\ee
where $\Lambda^{A}_{~B}$ represents the element of an arbitrary Lorentz transformation matrix which is position dependent and satisfies the relation $\eta_{AB}\Lambda^A_{~C}\Lambda^B_{~D}=\eta_{CD}$ at any spacetime point. In TG theory, the gravity is attributed to the spacetime torsion instead of the curvature. The torsion two form generally depends on both the tetrad and the spin connection,
\be\label{tor0}
\mathcal{T}^{A}_{~~\mu\nu}=2(\partial_{[\mu}e^{A}_{~\nu]}+\omega^{A}_{~B[\mu}e^{B}_{~\nu]})~,
\ee
 where the subscript brackets represent the antisymmetrization, so that the torsion tensor in spacetime is $\mathcal{T}^{\rho}_{~~\mu\nu}=e_A^{~\rho}\mathcal{T}^{A}_{~~\mu\nu}=2\hat{\Gamma}^{\rho}_{[\mu\nu]}$.
With these building blocks, the action of TG theory is written as
\bea\label{TGaction}
S_g=\frac{1}{2}\int d^4x ~{\|e\|}\mathbb{T}\equiv\int d^4x~{\|e\|} \left(-\frac{1}{2}\mathcal{T}_{\mu}\mathcal{T}^{\mu}+\frac{1}{8}\mathcal{T}_{\alpha\beta\mu}\mathcal{T}^{\alpha\beta\mu}
+\frac{1}{4}\mathcal{T}_{\alpha\beta\mu}\mathcal{T}^{\beta\alpha\mu}\right)~,
\eea
where ${ \|e\|}=\sqrt{-g}$ is the determinant of the tetrad, $\mathbb{T}$ is the torsion scalar, and $\mathcal{T}_{\mu}=\mathcal{T}^{\alpha}_{~~\mu\alpha}$ is the torsion vector.
This action is invariant under diffeomorphism and local Lorentz transformation.  Furthermore, it is identical to the Einstein-Hilbert action up to a surface term,
\be\label{graction}
S_g=\int d^4x \sqrt{-g}[-\frac{1}{2} R (e)-\nabla_{\mu}\mathcal{T}^{\mu}]~,
\ee
where the curvature scalar $R(e)$ is defined by the Levi-Civita connection and considered as being fully constructed from the metric, and in turn from the tetrad. The surface term does not affect the equation of motion and so TG theory can considered as a pure tetrad theory. The spin connection only contributes to the surface term, so it represents pure gauge in the TG action (\ref{TGaction}), and in practice we may fix a spin connection [as long as it satisfies the constraint of Eq. (\ref{omega})] and this imposes no effect on the equation of motion. The simplest choice is fixing to the Weitzenb\"{o}ck connection, $\omega^A_{~B\nu}=0$, which was frequently adopted in the literature.
With the Weitzenb\"{o}ck connection, the torsion two form is simply expressed as
\be\label{tor}
\mathcal{T}^{A}_{~~\mu\nu}=\partial_{\mu}e^{A}_{~\nu}-\partial_{\nu}e^{A}_{~\mu}~.
\ee
It deserves pointing out that in TG theory fixing a spin connection does not break the local Lorentz symmetry. One can prove that,  after taking the Weitzenb\"{o}ck connection, the action (\ref{TGaction}) is unchanged up to a surface term under the local Lorentz transformation, $e^{A}_{~\mu}\rightarrow \Lambda^{A}_{~B}(x)e^{B}_{~\mu}$.

The Weitzenb\"{o}ck connection has been continually used in many generalized or modified TG theories, such as the $f(\mathbb{T})$ models \cite{fT} where the torsion scalar in the
action (\ref{TGaction}) is replaced by its arbitrary function. In Ref. \cite{PVtele1} we have also adopted the Weitzenb\"{o}ck connection in our PV gravity model which modifies TG theory by adding a Nieh-Yan term in the action (\ref{TGaction}),
\be\label{NY1}
S_{NY}=\frac{c }{4}\int d^4x \sqrt{-g} ~\theta\,\mathcal{T}_{A\mu\nu}\widetilde{\mathcal{T}}^{A\mu\nu}~,
\ee
where $c$ is the coupling constant and $\widetilde{\mathcal{T}}^{A\mu\nu}=(1/2)\varepsilon^{\mu\nu\rho\sigma}\mathcal{T}^A_{~~\rho\sigma}$ is the dual of the torsion two form.
We also took into account the kinetic and potential terms of the scalar field and the action of other matter $S_m$ which coupled minimally through the metric (or the tetrad). The full action in its equivalent form is
\bea\label{oldmodel}
S=\int d^4x \sqrt{-g}\left[-\frac{R(e)}{2}
+\frac{c }{4}\,\theta\,\mathcal{T}_{A\mu\nu}\widetilde{\mathcal{T}}^{A\mu\nu}+\frac{1}{2}\nabla_{\mu}\theta\nabla^{\mu}\theta-V(\theta)\right]+S_m~.
\eea

Taking the Weitzenb\"{o}ck connection, as we have done in Ref. \cite{PVtele1}, the torsion two form $\mathcal{T}^{A}_{~~\mu\nu}$ is given by Eq. (\ref{tor}) and this gravity model is still a pure tetrad one.
The corresponding equations of motion of this model and its applications to the spatially flat FRW background universe and the cosmological perturbations around it are presented in detail in Ref. \cite{PVtele1}.
Here we only outline its main results: (i) the background evolution is the same as in GR, (ii) GWs present the velocity birefringence phenomenon, this is the direct effect of parity violation, (iii) the scalar field $\theta(x)$ is surprisingly not counted as an independent dynamical degree of freedom at
the linear-perturbation level.

We should point out here that  the PV term (\ref{NY1}) under the Weitzenb\"{o}ck condition breaks the local Lorentz symmetry explicitly.
Similar to many $f(\mathbb{T})$ models, when using the Weitzenb\"{o}ck connection, the PV gravity model (\ref{oldmodel}) is a pure tetrad theory, which generally does not have local Lorentz symmetry. This is because the change in the tetrad induced by a local Lorentz transformation cannot be compensated for by the change of the spin connection which is absent in the pure tetrad theory. The only exception is the TG theory, in which, as mentioned above, the change of the tetrad just brings a surface term into the action. Even though the explicit local Lorentz symmetry violation does not contradict with current experiments, it will make the conclusion less convincible because in practice the physical consequences obtained seem to rely on some special choices of local frames. For example, when we apply the model to cosmology, we should first assume some forms for the tetrad components; however, these forms are not uniquely determined by the metric as seen from the relation $g_{\mu\nu}=\eta_{AB}e^A_{~\mu}e^B_{~\nu}$. In other words, various (in fact infinite) choices of the tetrad forms will lead to the same metric. Before the calculation, we must make one specific choice on the tetrad---usually we choose the tetrads which have simple forms to simplify the subsequent calculations, this amounts to picking a special local frame out. Due to the lack of local Lorentz symmetry, different local frames which are related by local Lorentz transformations are not equivalent any more. One may doubt that the results from the calculations will depend on the specific choice of the local frame at the beginning, and cannot have general meanings. So it is questionable that one can do a more general analysis based on a particular choice of the tetrad form.

To avoid this problem, in this paper we give up the Weitzenb\"{o}ck condition in our model. In fact we will not impose any other constraint than Eq. (\ref{omega}) on the spin connection, this means we will restore the local Lorentz symmetry for our model (\ref{oldmodel}) by supplementing the eligible spin connections to the torsion so that it has the full expression  (\ref{tor0}). With the local Lorentz symmetry we can always choose the tetrad we prefer in calculations, because if at the beginning the tetrad at hand is not in the favored form we can obtain the tetrad what we want through a local Lorentz transformation. The price one must pay is that the spin connection is not fixed. The tetrads combined with spin connections are equivalent if they are connected by local Lorentz transformations.
Further in next sections, we will apply this new version of the model to cosmology in which the background universe has the general FRW spacetime, not just the spatially flat one. We want to see whether the conclusions drawn in Ref. \cite{PVtele1} are robust and want to know what will happen for the cases which haven't been studied in Ref. \cite{PVtele1}.

Now the torsion two form depends on the tetrad as well as the spin connection, so does the action (\ref{oldmodel}), which is a scalar in the spacetime as well as in the local internal space. Since the spin connection is constrained to be in the form of Eq. (\ref{omega}), we will use the Lorentz matrix element $\Lambda^{A}_{~B}$ to replace the spin connection $\omega^{A}_{~B\mu}$ as the fundamental variable in the action principle. So during the variation of the action, $\Lambda^{A}_{~B}$ and the tetrad $e^{A}_{~\mu}$ would be varied independently.
Furthermore, one can see that only the first-order derivatives of the fundamental variables appeared in the action, it is expected that this model is free from the ghost instability caused by higher-order derivatives.

The model  has two kinds of gauge symmetries---the diffeomorphism invariance and the local Lorentz invariance---the latter transformation makes the following change
\be\label{LT}
e^{A}_{~\mu}\rightarrow(L^{-1})^{A}_{~B}e^{B}_{~\mu}~,~ \Lambda^{A}_{~B}\rightarrow\Lambda^{A}_{~C}L^{C}_{~B}~,
\ee
where $L^{A}_{~B}(x)$ is also the element of Lorentz matrix. We would like to use different notations to distinguish two kinds of Lorentz matrices, $\Lambda^{A}_{~B}(x)$ is used to express the spin connection as in Eq. (\ref{omega}), but $L^{A}_{~B}(x)$ represents the local transformation which makes a shift from one local frame to another. It's easy to prove that the metric $g_{\mu\nu}$ and torsion tensor $\mathcal{T}_{~\mu \nu}^{\rho}$
are invariant under the local Lorentz transformation (\ref{LT}), so is the action (\ref{oldmodel}).

The equations of motion follow from the variation of the action (\ref{oldmodel}) with respect to $e^{A}_{~\mu}$ and $\Lambda^{A}_{~B}$ separately
\bea
 G^{\mu\nu}+N^{\mu\nu}&=&T^{\mu\nu}+T^{\mu\nu}_{\theta}~,\label{eom1}\\
 N^{[\mu\nu]}&=&0~,\label{eom2}
\eea
where $G^{\mu\nu}$ is the Einstein tensor, $T^{\mu\nu}=-(2/\sqrt{-g})(\delta S_m/\delta g_{\mu\nu})$
and $T^{\mu\nu}_{\theta}=[V(\theta)-\nabla_{\alpha}\theta\nabla^{\alpha}\theta/2]g^{\mu\nu}+\nabla^{\mu}\theta\nabla^{\nu}\theta$
are the energy-momentum tensors for the matter and the scalar field $\theta$ respectively, and $N^{\mu \nu}=c\, e_{A}^{~\,\nu} \partial_{\rho} \theta\, \widetilde{\mathcal{T}}^{A \mu \rho}$.
These equations with same forms had also been obtained in Ref. \cite{PVtele1}, but there are two differences needed to be pointed out. First, in Ref. \cite{PVtele1} the torsion $\mathcal{T}^{A}_{~~\mu\nu}$ contained in the tensor $N^{\mu\nu}$ was only determined by the tetrad, as expressed in Eq. (\ref{tor}), but here it depends on both of them and has the full expression (\ref{tor0}); Second, in Ref. \cite{PVtele1} the second equation (\ref{eom2}) was not obtained from the action principle but deduced from the first equation (\ref{eom1}) that requires all the tensors appeared in it should be symmetric under the permutation of indices $\mu$ and $\nu$, but here it is obtained from the variation of the action with respect to $\Lambda^{A}_{~B}$, and it is certainly consistent with Eq. (\ref{eom1}).

From another viewpoint, the consistency showed that the equation of motion (\ref{eom2}) from the variation of $\Lambda^{A}_{~B}$ is not independent of Eq. (\ref{eom1}), it is just the antisymmetric part of the latter.
This subtlety can be explained as follows. First, the local Lorentz transformation (\ref{LT}) can make $\Lambda^{A}_{~B}$ go through all possible values, because all Lorentz matrices form a group.
In consequence, no matter what change $\Lambda^{A}_{~B}$ undergoes, it can change back through a local Lorentz transformation.
That is to say, for a local Lorentz invariant quantity $F=F(e, \Lambda)$,
any change in $F$ caused only by a change in $\Lambda^{A}_{~B}$ can be equivalent to a change in $F$ caused only by a change in $e^{A}_{~\mu}$.
This is pictured as
$$
F(e,\Lambda+\delta\Lambda)\overset{(\ref{LT})}{=} F(e+\delta e, \Lambda)~~ \text{for}~\forall \delta\Lambda~.
$$
But its inverse statement does not hold because the local Lorentz transformation cannot make $e^{A}_{~\mu}$ to go through all possible values.
The action $S$ of our model is invariant under the local Lorentz transformation (\ref{LT}), therefore any $\delta S$ caused by $\delta\Lambda^{A}_{~B}$ can always be
equal to that due to $\delta e^{A}_{~\mu}$. It means that from the viewpoint of variational principle,
requiring $S$ to take the extremum under $\delta e^{A}_{~\mu}$  already includes the case where $S$ takes the extremum under $\delta\Lambda^{A}_{~B}$.
Hence, the equations obtained for the variations of $e^{A}_{~\mu}$  already contains the equations obtained for the variation of $\Lambda^{A}_{~B}$.

Compared with GR, the model has 16+6 basic variables of gravity, while GR has only ten.
At the same time, the model has six more constraint equations and 6 local Lorentz gauge symmetries.
Since both theories are diffeomorphism invariant,
it can be expected that our model will have as many physical dynamical degrees of freedom as GR.

In addition, the Bianchi identity $\nabla_{\mu}G^{\mu\nu}=0$ and the covariant conservation law, $\nabla_{\mu} T^{\mu \nu}=0$, demand that $\nabla_{\mu} N^{\mu \nu}=\nabla_{\mu} T_{\theta}^{\mu \nu}$.
This brings no further constraint, in stead it is consistent with the Klein-Gordon equation
\be\label{eom3}
\Box\theta+V_{\theta}-\frac{c}{4} \mathcal{T}_{A\mu\nu}\widetilde{\mathcal{T}}^{A\mu\nu}=0~,
\ee
which is obtained from the variation of the action (\ref{oldmodel}) with respect to the scalar field $\theta$.
Here and in the following we will use $V_{\theta}$ to denote the first derivative of the potential to the scalar field.

Finally, we briefly explain why we introduce the kinetic and potential terms of the scalar field $\theta$ into the action. Without them, the dynamics and the energy momentum tensor of $\theta$ will be absent, it can be seen from (\ref{eom1}) that
the Bianchi identities and conservation of energy momentum tensor of matter impose an additional constraint $\nabla_{\mu} N^{\mu \nu}=0$.
This extra constraint together with the constraint (\ref{eom2}) put too strong restrictions on the space of solutions, which makes some common solutions in GR nonexistent.
This is similar to the case of nondynamical CS gravity \cite{Jackiw:2003pm,Alexander:2009tp}.
Later in this paper we will see that if the dynamics of $\theta$ are not turned on, there will be no real closed FRW solution.

\section{Cosmology background solution}\label{cosmology background}

Now we apply the new version of our model with full local Lorentz covariance to cosmology. In this section, we only consider the background evolutions and leave the discussions on the cosmological perturbations to next sections.

In Ref.\,\cite{PVtele1} we only considered the spatially flat FRW background, and assumed that the tetrad was diagonal in form.
Here we will discuss the more general FRW background, the metric has the following well-known form in spherical coordinate system
\be\label{FRWmetric}
ds^{2}=g_{\mu\nu}dx^{\mu}dx^{\nu}=a(\eta)\left(d\eta^{2}-\frac{dr^{2}}{1-K r^{2}}-r^{2}d\Omega^{2}\right)~,
\ee
where $a(\eta)$ is the scale factor of the universe, $\eta$ is the conformal time and
$d\Omega^{2}=d\varphi^{2}+(\sin\varphi)^{2} d\phi^{2}$ is the metric on the unit sphere. We have used $\varphi$ and $\phi$ to represent the polar angle and azimuth angle respectively.
The subspace of $\eta={\rm constant}$ is homogeneous and isotropic, so there are six Killing vector fields $\hat{\xi}$ in total.
When applying our model to this spacetime and for the convenience of looking for solutions, we consider the affine connection which is homogeneous and isotropic,
\be\label{condition}
\mathcal{L}_{\hat{\xi}}\hat{\Gamma}^{\rho}_{~\mu\nu}=0~,
\ee
where $\mathcal{L}_{\hat{\xi}}$ is the Lie derivative along the Killing vector field $\hat{\xi}$. This condition has been used in TG theory \cite{telesymmetry1,telesymmetry2}. One may find solutions without the above requirement on the connection, we leave this for future work. As mentioned before, the affine connection relates to the spin connection via the equation $\hat{\Gamma}^{\rho}_{~\mu\nu}= e_{A}^{~\,\rho}\partial_{\mu}e^{A}_{~\nu}+ e_{A}^{~\,\rho}e^{B}_{~\nu}\omega^{A}_{~B\mu}$. Although $\hat{\Gamma}^{\rho}_{~\mu\nu}$ is coordinate dependent, the Lie derivative of $\hat{\Gamma}^{\rho}_{~\mu\nu}$ does not depend on the coordinate. Hence the condition (\ref{condition}) is unambiguous.
There is a benefit to requesting this condition (\ref{condition}),
for any tensor $T$ constructed from the metric, torsion and their covariant derivatives associated with the Levi-Civita connection or the affine connection,
it always has the properties of homogeneity and isotropy, i.e., $\mathcal{L}_{\hat{\xi}}T=0$.

The general solutions to Eqs.~(\ref{FRWmetric}) and (\ref{condition}) have been discussed in Ref.\,\cite{telesymmetry1,telesymmetry2}
within the framework of TG. Below we briefly introduce the solution and its application to our model.
In this paper, we only consider the case where both $e^{A}_{~\mu}$ and $\Lambda^{A}_{~B}$ are real numbers.

\subsection{Spatially flat FRW universe}

In the case of spatially flat FRW universe, $K=0$, one of the solutions can be parametrized as
\be\label{flat}
{e}^{A}_{~\mu}=a(\eta)\left( \begin{array}{cccc}
                      1 & 0 & 0 & 0\\
                      0 & \sin\varphi\cos\phi & r\cos\varphi\cos\phi & -r\sin\varphi\sin\phi\\
                      0 & \sin\varphi\sin\phi & r\cos\varphi\sin\phi & r\sin\varphi\cos\phi\\
                      0 & \cos\varphi & -r\sin\varphi & 0
               \end{array}  \right)~,
               ~~~\Lambda=\mathring{\Lambda}~.
\ee
Where $\mathring{\Lambda}$ is a global Lorentz matrix, which does not depend on spacetime.
But no matter which global Lorentz matrix is taken, it will not affect any result discussed below.
So the specific expression of $\mathring{\Lambda}$ does not matter and it is not necessary to write its concrete form out explicitly.

All other solutions under the same constraints on the metric and affine connection differ from the solution (\ref{flat}) only by local Lorentz transformations (\ref{LT}).
The local Lorentz symmetry is a gauge symmetry in our model.
So physically, solution (\ref{flat}) is the only solution when $K=0$.
It can be verified that solution (\ref{flat}) satisfies the constraint equation (\ref{eom2}).
Putting Eq.~(\ref{flat}) into Eqs.~(\ref{eom1}) and (\ref{eom3}), we can get
\bea
& & 3 \mathcal{H}^{2}=a^{2}\left(\rho_{\theta}+\rho\right)~,\\
& & 2 \mathcal{H}^{\prime}+\mathcal{H}^{2}=-a^{2}\left(p_{\theta}+p\right)~,\\
& & \theta^{\prime \prime}+2 \mathcal{H} \theta^{\prime}+a^{2} V_{\theta}=0~,
\eea
where prime represents the derivative with respect to the conformal time, $\mathcal{H}=a'/a$ is the conformal Hubble
rate, $\rho_{\theta}=\theta^{\prime 2} /\left(2 a^{2}\right)+V$ and $p_{\theta}=\theta^{\prime 2} /\left(2 a^{2}\right)-V$
are the energy density and pressure of the $\theta$ field, and $\rho$ and $p$ denote
the energy density and pressure of other matter. All of them only depend on time.
The background equations are exactly the same as in GR. This means the PV term has no effect on the flat FRW background. This result is the same as that of Ref. \cite{PVtele1}.

\subsection{Closed FRW universe}

In the case of closed FRW universe, $K>0$, the solution is parametrized as
\be\label{closed}
e^{A}_{~\mu}=a(\eta)\left( \begin{array}{cccc}
                      1 & 0 & 0 & 0 \\
                      0 & \sin\varphi \cos\phi/\mathcal{R} & r\mathcal{R}_{1} & r\sin\varphi\mathcal{R}_{2}\\
                      0 & \sin\varphi \sin\phi/\mathcal{R} & r\mathcal{R}_{3} & r\sin\varphi\mathcal{R}_{4}\\
                      0 & \cos\varphi/\mathcal{R} & -r\mathcal{R}\sin\varphi & -\mathcal{K}r^{2}(\sin\varphi)^{2}
               \end{array}  \right),
               ~\Lambda=\mathring{\Lambda}~.
\ee
Here $\mathcal{R}=\sqrt{1-Kr^{2}}$, $\mathcal{R}_{1}= \mathcal{R} \cos\varphi\cos\phi+\mathcal{K}r\sin\phi$,
$ \mathcal{R}_{2}=\mathcal{K}r\cos\varphi\cos\phi-\mathcal{R}\sin\phi$,
$ \mathcal{R}_{3}=\mathcal{R}\cos\varphi\sin\phi-\mathcal{K}r\cos\phi$,
$ \mathcal{R}_{4}=\mathcal{R}\cos\phi+\mathcal{K}r\cos\varphi\sin\phi$,
and the real parameter $\mathcal{K}$ satisfies $\mathcal{K}^{2}=K$.
One can see that when $K\rightarrow0$, the Eq.~(\ref{closed}) returns to (\ref{flat}).
All other solutions differ from Eq.~(\ref{closed}) only by local Lorentz transformations (\ref{LT}).
So physically Eq.~(\ref{closed}) is the only solution when $K>0$.
It can be verified that this solution (\ref{closed}) satisfies the constraint equation (\ref{eom2}).
Substitute this solution into the Eqs.~(\ref{eom1}) and (\ref{eom3}), one obtains the background equations,
\bea
& & 3 (\mathcal{H}^{2}+\mathcal{K}^{2})=a^{2}\left(\rho_{\theta}+\rho\right)~,\\
& & 2 \mathcal{H}^{\prime}+\mathcal{H}^{2}+\mathcal{K}^{2}-2c\mathcal{K}\theta'=-a^{2}\left(p_{\theta}+p\right)~,\\
& &\theta^{\prime \prime}+2 \mathcal{H} \theta^{\prime}+a^{2} V_{\theta}-6c\mathcal{K}\mathcal{H}=0~.
\eea
These modified corresponding equations in GR and showed the effects of the PV term on the closed FRW background.

Now, we can see that if the scalar $\theta$ is not a dynamical field, its kinetic and potential terms are absent in the action at the beginning,  an additional constraint $\nabla_{\mu} N^{\mu \nu}=0$ emerged,
but this is violated by the solution (\ref{closed}). So if $\theta(x)$ is not a dynamical field, we are not able to have a closed FRW solution under Eq. (\ref{condition}) in our model.

\subsection{Open FRW universe}

In the case of open FRW universe, $K<0$, the solution is parametrized as
\be\label{open}
{e}^{A}_{~\mu}=a(\eta)\left( \begin{array}{cccc}
                      1 & 0 & 0 & 0\\
                      0 & 1/\mathcal{R} & 0 & 0\\
                      0 & 0 & r & 0\\
                      0 & 0 & 0 & r\sin\varphi
               \end{array}  \right)~,~~~
\Lambda=\mathring{\Lambda}\cdot\left(\begin{array}{cccc}
                               \mathcal{R}&\mathcal{K}r&0&0\\
                               \mathcal{K}r\sin\varphi\cos\phi & \mathcal{R}\sin\varphi\cos\phi & \cos\varphi\cos\phi & -\sin\phi\\
                               \mathcal{K}r\sin\varphi\sin\phi & \mathcal{R}\sin\varphi\sin\phi & \cos\varphi\sin\phi & \cos\phi\\
                               \mathcal{K}r\cos\varphi & \mathcal{R}\cos\varphi & -\sin\varphi & 0
                               \end{array}\right)
\ee
Here $\mathcal{R}=\sqrt{1-Kr^{2}}$ and the real parameter $\mathcal{K}$ satisfies $\mathcal{K}^{2}=-K$.
It can be proved that when $K\rightarrow0$, the solution (\ref{open}) also returns to (\ref{flat}) through a local Lorentz transformation (\ref{LT}).
All the other solutions differ from solution (\ref{open}) merely by local Lorentz transformations (\ref{LT}).
So physically Eq.~(\ref{open}) is the only solution when $K<0$.
It can be verified that this solution (\ref{open}) satisfies the constraint Eq.~(\ref{eom2}).
Substitute it into the Eqs.~(\ref{eom1}) and (\ref{eom3}), one obtains the background equations,
\bea
& & 3 (\mathcal{H}^{2}-\mathcal{K}^{2})=a^{2}\left(\rho_{\theta}+\rho\right)~,\\
& & 2 \mathcal{H}^{\prime}+\mathcal{H}^{2}-\mathcal{K}^{2}=-a^{2}\left(p_{\theta}+p\right)~,\\
& & \theta^{\prime \prime}+2 \mathcal{H} \theta^{\prime}+a^{2} V_{\theta}=0~.
\eea
These equations are the same as those in GR, so the PV term has no effect on the open FRW background.

By the way, we can indeed do a local Lorentz transformation on Eq.~(\ref{open}) to make $\Lambda=\mathring{\Lambda}$, just like the cases of $K=0$ and $K>0$.
But this will make $e^{0}_{~i}\neq0~,~e^{a}_{~0}\neq0$ and make it inconvenient to perform decomposition of the tetrad perturbation which will be discussed in the next section.
So we use Eq.~(\ref{open}) instead of the equivalent solution satisfying $\Lambda=\mathring{\Lambda}$.

The above equations of the background evolutions are consistent with the results of Ref. \cite{Hohmann:2020dgy}, where the tetrad field was parametrized separately as the ``vector branch" and ``axial branch" \{see the Eqs.~(29) and (30) in \cite{Hohmann:2020dgy}\}, and these two branches match in the case of a spatially-flat universe ($K=0$). If we keep all the components of the tetrad field and all the terms in the equations real, as we required in this paper, one can only consider the vector branch for the open universe ($K<0$) and the axial branch for the closed universe ($K>0$). It was shown in Ref. \cite{Hohmann:2020dgy} (Sec. IV. B) that the PV modifications bring changes only for the axial branch, so they can only have effects on the background evolutions of the closed universe.

\subsection{Additional comments}

This subsection is mainly prepared for the following section and for declaring of some conventions.
The FRW metric can always be expressed as
\be
ds^{2}=a^{2}(d\eta^{2}-\gamma_{ij}dx^{i}dx^{j})~,
\ee
where $\gamma_{ij}$ can be regarded as the metric of three-dimensional hypersurface, and its inverse is denoted as $\gamma^{ij}$.
The corresponding spatial covariant derivative will be denoted by $D_{i}$.
From (\ref{flat}), (\ref{closed}) and (\ref{open}), we can see that ${e}^{A}_{~\mu}$ can  always
be expressed as
\be\label{tetradsimple}
{e}^{0}_{~0}=a~,~ {e}^{0}_{~i}=0~,~{e}^{a}_{~0}=0~,~{e}^{a}_{~i}=a\gamma^{a}_{~i}~,
\ee
where $\gamma^{a}_{~i}$ can be regarded as the spatial tetrad on the three-dimensional hypersurface
(sometimes it is  called dreibein).
Its dual tetrad is denoted as $\gamma_{a}^{~i}$.
From $g_{\mu \nu}=\eta_{A B} e_{~\mu}^{A} e_{~\nu}^{B}$, we have $\gamma_{ij}=\delta_{ab}\gamma^{a}_{~i}\gamma^{b}_{~j}$
and therefore $\gamma^{ij}=\delta^{ab}\gamma_{a}^{~i}\gamma_{b}^{~j}$.
We also denote $\varepsilon_{ijk}$ as  spatial volume element compatible with the spatial metric $\gamma_{ij}$.

In the rest of this paper, the spatial coordinate indices ``i, j, ..." of variables will be raised and lowered by $\gamma^{ij}$ and $\gamma_{ij}$ and the spatial internal indices ``a, b, ..." will be raised and lowered by $\delta^{ab}$ and $\delta_{ab}$.
For example, $\gamma^{ai}=\gamma^{a}_{~j}\gamma^{ji}=\delta^{ab}\gamma_{b}^{~i}$.
And all the transformations between these two kinds of indices
will be performed  by $\gamma^{a}_{~i}$ and $\gamma_{a}^{~i}$.
For example, $\omega^{i}_{~jk}=\gamma_{a}^{~i}\gamma^{b}_{~j}\omega^{a}_{~bk}$.

Here, we would like to stress again the advantage of the model with full local Lorentz covariance. As mentioned before, given a metric the form for the tetrad is not uniquely determined. For FRW universe, we have infinite choices on the tetrad forms, all of them lead to the same metric (\ref{FRWmetric}).  We just mentioned that we can always take the simple diagonal form of Eq. (\ref{tetradsimple}) for the tetrad. This is because, if it is not so at the beginning, we can always get the form of Eq. (\ref{tetradsimple}) through a local Lorentz transformation which is now the gauge symmetry of the model.
Due to this advantage, we can also safely take simple ansatz for the perturbed tetrad components, this is what we will do in next sections.

\section{perturbations and gauge transformations}\label{perturbation}

With the FRW background solutions obtained in previous section, we are ready to study the linear cosmological perturbations around the FRW background. In this section we will focus on the scalar-vector-tensor (SVT) decomposition and the gauge transformations of the cosmological perturbations.

\subsection{Perturbations and SVT decomposition}

The tetrad perturbations in cosmology have been studied extensively in the literature, such as \cite{tetradper1,tetradper2}.
We will also consider the cosmological perturbations of the spin connection for a consistent and complete study.
Here we briefly introduce how to do the SVT decomposition of perturbations to the tetrad and Lorentz matrices, taking the general FRW universe as the background.

A spatial covariant vector $\mathcal{V}_{i}$ can always be decomposed into
\be
\mathcal{V}_{i}=D_{i}\mathcal{V}+\mathcal{V}^{V}_{i}~.
\ee
The superscript $V$ represents that it satisfies the transverse condition $D^{i}\mathcal{V}^{V}_{i}=0$.
Because a spatial covariant antisymmetric tensor field $\mathcal{A}_{ij}$ can always be Hodge-dual to a covariant vector
$\mathcal{V}_{i}=(1/2)\varepsilon_{ijk}\mathcal{A}^{jk}$, a spatial covariant antisymmetric tensor $\mathcal{A}_{ij}$ can be decomposed into
\be
\mathcal{A}_{ij}=\varepsilon_{ij}^{~~k}(D_{k}\mathcal{V}+\mathcal{V}^{V}_{k})~.
\ee
A symmetric spatial tensor $S_{ij}$ can always be decomposed as
\be
S_{ij}=\hat{S}\gamma_{ij}+D_{i}D_{j}S+D_{(i}S^{V}_{j)}+S^{T}_{ij}~,
\ee
where the subscript parentheses denotes the symmetrization,
 and the superscript $T$ means traceless and transverse.
Finally, we know that any second-rank tensor $T_{ij}$ can always be expressed as the sum of a symmetric tensor $S_{ij}$ and an antisymmetric tensor $\mathcal{A}_{ij}$,
so it can be decomposed as
\be
T_{ij}=S_{ij}+\mathcal{A}_{ij}=\hat{S}\gamma_{ij}+D_{i}D_{j}S+D_{i}S^{V}_{j}+\varepsilon_{ij}^{~\,k}(D_{i}\mathcal{V}+\mathcal{V}^{V}_{i})+S^{T}_{ij}~.
\ee
Note that in the last expression, we have absorbed the antisymmetric part of $D_{i}S^{V}_{j}$ into $\mathcal{V}$ and $\mathcal{V}^{V}_{i}$ in terms of the method of Hodge duality:
$ D_{[i}S^{V}_{j]}=\varepsilon_{ij}^{~~k}(D_{k}\mathcal{U}+\mathcal{U}^{ V}_{k})$.

\subsubsection{Tetrad perturbations}

Note that $e^{0}_{~0}$ can be regarded as spatial scalar field,
$e^{0}_{~i}$ and $\gamma_{ij}\gamma_{a}^{~j}e^{a}_{~0}$ can be regarded as spatial covariant vector fields,
and $\gamma_{jk}\gamma_{a}^{~k}e^{a}_{~i}$ can be regarded as spatial covariant second-rank tensor field.
Hence, all the tetrad components can be written as
\bea\label{tetrad1}
& & e^{0}_{~0}=a(1+A)~,~e^{0}_{~i}=a(D_{i}\beta+\beta^{V}_{~i})~,~e^{a}_{~0}=a\gamma^{ai}(D_{i}\chi+\chi^{v}_{i})~,\nonumber\\
& & e^{a}_{~i}=a\gamma^{aj}\left[(1-\psi)\gamma_{ij}+D_{i}D_{j}\alpha+D_{i}\alpha^{V}_{j}-\varepsilon_{ij}^{~\,k}
(D_{k}\lambda+\lambda^{V}_{k})+\frac{1}{2}h^{T}_{ij}\right]~,
\eea
so that the perturbed metric components have the following standard forms
\bea\label{metric1}
& &g_{00}=a^{2}(1+2A)~,~ g_{0i}=-a^{2}(D_{i}B+B^{V}_{i})~,\nonumber\\
& &g_{ij}=-a^{2}\left[(1-2\psi)\gamma_{ij}+2D_{i}D_{j}\alpha+D_{i}\alpha_{j}^{V}+D_{j}\alpha_{i}^{V}+h^{T}_{ij}\right]~,
\eea
where $B=\chi-\beta$ and $B^{V}_{i}=\chi^{V}_{i}-\beta^{V}_{i}$.
Besides the familiar scalar perturbations ($A, \chi-\beta, \psi, \alpha$), vector perturbations ($\chi_i^V-\beta_i^V, \alpha^V_i$), and tensor perturbation ($h^T_{ij}$) in the metric, the parametrization of tetrad brings six extra variables which are scalar perturbation $\lambda, \chi+\beta$ and vector perturbation $\lambda_i^V, \chi_i^V+\beta_i^V$.

Again, we have assumed the forms in Eq. (\ref{tetrad1}) for the tetrad components in the perturbed universe. This ansatz leads to the metric (\ref{metric1}). But the tetrad can have other forms leading to the same metric (\ref{metric1}). We do not need to worry about the special status of the tetrad we have chosen in Eq. (\ref{tetrad1}). Because the model has the local Lorentz symmetry, if at the beginning we have a tetrad different from that in Eq. (\ref{tetrad1}), we can always shift it to the one in Eq. (\ref{tetrad1}) by a local Lorentz transformation.

\subsubsection{Perturbations to Lorentz matrix elements}

As mentioned before, in the covariant version of our model we can always take simple forms for the tetrad, but there is a price that the spin connection or the $\Lambda$-matrix element cannot be fixed. In the cosmological perturbation theory one should also consider the perturbation to the Lorentz matrix elements. In the following we denote the unperturbed Lorentz matrix element as $\bar{\Lambda}^{A}_{~B}$, and the perturbed one as $\Lambda^{A}_{~B}$.
Since the Lorentz matrices form a group, $(\bar{\Lambda}^{-1})^{A}_{~C}\Lambda^{C}_{~B}$ is also a Lorentz matrix element.
Thus $(\bar{\Lambda}^{-1})^{A}_{~C}\Lambda^{C}_{~B}$ can be exponentialized as $\exp(\epsilon)^{A}_{~B}$,
where $\epsilon_{AB}=\eta_{AC}\epsilon^{C}_{~B}$  is an antisymmetric matrix and has six independent components.
At the linear perturbation level, the difference between $\bar{\Lambda}^{A}_{~B}$  and $\Lambda^{A}_{~B}$ is very small,
so $\epsilon_{AB}$ is a small quantity.
Thus the perturbed Lorentz matrix can be parametrized as
\be
\Lambda=\bar{\Lambda}\exp(\epsilon)~.
\ee
Note that $\gamma^{a}_{~i}\epsilon^{0}_{~a}$ can be regarded as a spatial covariant vector,
and $\gamma_{ai}\gamma^{b}_{~j}\epsilon^{a}_{~b}$ can be regarded as an antisymmetric spatial covariant tensor,
 $\epsilon$ can be decomposed as
\be
\epsilon^{0}_{~a}=\gamma_{a}^{~i}(D_{i}\kappa+\kappa^{V}_{i})~,
~\epsilon^{a}_{~b}=\gamma^{ai}\gamma_{b}^{~j}\varepsilon_{ij}^{~\,k}(D_{k}\tau+\tau^{V}_{k})~.
\ee
According to this method, the total six independent perturbative degrees of freedom of the local Lorentz matrices can be decomposed
into two scalar components $\kappa, \tau$ and four vector components $\kappa^{V}_{i}, \tau^{V}_{i}$. There is no tensor component appeared in this decomposition.

\subsection{Gauge transformations}

In this subsection, we discuss how the above perturbative variables change under the infinitesimal gauge transformations.
Our model has two kinds of gauge symmetries: Diffeomorphism invariance and local Lorentz symmetry.

\subsubsection{Infinitesimal diffeomorphism}
The diffeomorphism induced by the infinitesimal vector field $\xi^{\mu}$ leads following changes for the tetrad and local Lorentz matrix
$\delta e^{A}_{~\mu}=-\mathcal{L}_{\xi}e^{A}_{~\mu}$ and $\delta \Lambda^{A}_{~B}=-\mathcal{L}_{\xi}\Lambda^{A}_{~B}$.
The vector field $\xi^{\mu}$ can be further decomposed into two scalar degrees of freedom $\xi^{0}, \xi$ and  two vector degrees of freedom $\xi^{V}_{i}$,
where $\xi^{\mu}=(\xi^{0}, \xi^{i}$), $\xi_{i}=\gamma_{ij}\xi^{j}=D_{i}\xi+\xi^{V}_{i}$.
For convenience, we also define $\tilde{\xi}_{i}=(1/2)\varepsilon_{ik}^{~~l}\gamma_{a}^{~k}\gamma^{b}_{~l}\omega^{a}_{~bj}\xi^{j}$,
 and decompose it into $\tilde{\xi}_{i}=D_{i}\tilde{\xi}+\tilde{\xi}^{V}_{i}$.
Through calculations, it is not difficult to obtain the transformation laws of the perturbations under the infinitesimal diffeomorphism
\bea\label{trans1}
& &\nonumber A\rightarrow A-{\xi^{0}}'-\mathcal{H}\xi^{0}~,~\psi\rightarrow\psi+\mathcal{H}\xi^{0}~,~\beta\rightarrow\beta-\xi^{0}~,~\chi\rightarrow\chi-\xi'~,\\
& &\nonumber \alpha\rightarrow\alpha-\xi~,~\beta^{V}_{i}\rightarrow\beta^{V}_{i}~,~\chi^{V}_{i}\rightarrow\chi^{V}_{i}-{\xi^{V}_{i}}'
~,~\alpha^{V}_{i}\rightarrow\alpha^{V}_{i}-\xi^{V}_{i}~,~h^{T}_{ij}\rightarrow h^{T}_{ij}~,\\
& &\nonumber \kappa\rightarrow  {\left\{\begin{array}{ll}
                                \kappa & (K\geq0)\\
                                \kappa-\mathcal{K}\xi & (K<0)
                              \end{array}\right.}~,~
            \kappa^{V}_{i}\rightarrow{\left\{\begin{array}{ll}
                                \kappa^{V}_{i} & (K\geq0)\\
                                \kappa^{V}_{i}-\mathcal{K}\xi^{V}_{i} & (K<0)
                              \end{array}\right.}~,~
                              \tau\rightarrow{\left\{\begin{array}{ll}
                                \tau & (K\geq0)\\
                                \tau-\tilde{\xi} & (K<0)
                              \end{array}\right.}~,\\
 & &           \tau^{V}_{i}\rightarrow{\left\{\begin{array}{ll}
                                \tau^{V}_{i} & (K\geq0)\\
                                \tau^{V}_{i}-\tilde{\xi}^{V}_{i} & (K<0)
                              \end{array}\right.}~,~
 \lambda\rightarrow{\left\{\begin{array}{ll}
                                \lambda-\mathcal{K}\xi & (K\geq0)\\
                                \lambda+\tilde{\xi} & (K<0)
                              \end{array}\right.}~,
 \lambda^{V}_{i}\rightarrow{\left\{\begin{array}{ll}
                                \lambda^{V}_{i}-\mathcal{K}\xi^{V}_{i} & (K\geq0)\\
                                \lambda^{V}_{i}+\tilde{\xi}^{V}_{i} & (K<0)
                              \end{array}\right.}~.
\eea
By the way, the scalar field $\theta$ can be decomposed into $\theta(\eta, \vec{x})=\theta(\eta)+\delta \theta(\eta,\vec{x})$, its fluctuation is not diffeomorphism invariant, i.e., $\delta\theta\rightarrow\delta\theta-\theta'\xi^{0}$.

\subsubsection{Infinitesimal Local Lorentz Transformation}

The infinitesimal Lorentz transformation (\ref{LT}) can be expressed as $L^{A}_{~B}=\delta^{A}_{~B}+\delta L^{A}_{~B}$, and it is easy to see that  $\delta L_{AB}=\eta_{AC}\delta L^{C}_{~B}=-\delta L_{BA}$ is antisymmetric.
Similar to the previous analyses, $\delta L_{AB}$ can be decomposed as
\be
\delta L^{0}_{~a}=\gamma_{a}^{~i}(D_{i}\ell+\ell^{V}_{i})~,
~\delta L^{a}_{~b}=\gamma^{ai}\gamma_{b}^{~j}\varepsilon_{ij}^{~\,k}(D_{i}l+l^{V}_{i})~.
\ee
The transformation laws of the perturbations under the infinitesimal local Lorentz transformation can be obtained in terms of Eq. (\ref{LT}):
\bea\label{trans2}
& &\nonumber A\rightarrow A~,~\beta\rightarrow\beta-\ell~,~\chi\rightarrow\chi-\ell~,~
\lambda\rightarrow\lambda-l~,~\psi\rightarrow\psi~,~\alpha\rightarrow\alpha~,~\kappa\rightarrow\kappa+\ell~,~\tau\rightarrow\tau+l~,~h^{T}_{ij}\rightarrow h^{T}_{ij}~,\\
& & \beta^{V}_{i}\rightarrow\beta^{V}_{i}-\ell^{V}_{i}~,~\chi^{V}_{i}\rightarrow\chi^{V}_{i}-\ell^{V}_{i}~,~\lambda^{V}_{i}\rightarrow\lambda^{V}_{i}-l^{V}_{i}~,~
\alpha^{V}_{i}\rightarrow\alpha^{V}_{i}~,~\kappa^{V}_{i}\rightarrow\kappa^{V}_{i}+\ell^{V}_{i}~,~
\tau^{V}_{i}\rightarrow\tau^{V}_{i}+l^{V}_{i}~.
\eea
One can see that all perturbations which enter the metric, such as $A, \psi, B=\chi-\beta, \alpha, B^{V}_{i}=\chi^{V}_{i}-\beta^{V}_{i}, \alpha^{V}_{i}, h^{T}_{ij}$,  are local Lorentz invariant.

\subsubsection{A special gauge}

In this subsection, we consider a specific gauge which will be fixed in the next section to simplify the calculations.
Firstly, from (\ref{trans1}), we can always choose $\xi$ and $\xi^{V}_{i}$ to make $\alpha=0, \alpha^{V}_{i}=0$.
Secondly, from (\ref{trans2}), we choose $\ell$, $l$, $\ell^{V}_{i}$ and $l^{V}_{i}$ to make $\kappa=0, \tau=0, \kappa^{V}_{i}=0, \tau^{V}_{i}=0$.
In this process, the conditions $\alpha=0, \alpha^{V}_{i}=0$ do not change.
Finally we adjust $\xi^{0}$ to make $\delta\theta=0$.
In this process, the conditions $\alpha=0,~\alpha^{V}_{i}=0,~\kappa=0,~\tau=0,~\kappa^{V}_{i}=0,~\tau^{V}_{i}=0$ do not change.
Therefore, we can always go to the gauge: $\delta\theta=0,~\alpha=0,~\alpha^{V}_{i}=0,~\kappa=0,~\tau=0,~\kappa^{V}_{i}=0,~\tau^{V}_{i}=0$.
In next section, we will do calculations under this gauge.

\section{quadratic actions for scalar and tensor perturbations}\label{quadratic action}

When applying our PV gravity model to the early universe, such as the inflationary epoch, we should attach much importance to the quantum origins of primordial perturbations.
For this purpose, one need to quanitze the cosmological perturbations, and during which the quadratic actions of perturbations are indispensable. In this section, we will discuss the quadratic actions for perturbations in our model.
For convenience, we ignore other mater, so that $S_m=0$.

The direct calculations show that in the quadratic action, scalar perturbations, vector perturbations and tensor perturbations are decoupled.
The vector perturbations are not dynamical degrees of freedom, this is the same with the case in GR. Therefore we will neglect vector perturbations in the rest of this paper and only discuss the quadratic actions of scalar and tensor perturbations.

In the following calculations, we will take the gauge $\delta\theta=0,~\alpha=0,~\kappa=0,~\tau=0$.
We also introduce the gauge invariant variable $\zeta=-(\psi+\mathcal{H}\delta\theta/\theta')$
which is the curvature perturbation of the hypersurfaces with constant $\theta$ field.
Then, we can choose $A$, $\zeta$, $B$, $\beta$, $\lambda$ and $h^{T}_{ij}$ as independent variables. Under this gauge condition, the dynamics of the scalar field at the linear level are totally shifted to the gauge invariant quantity $\zeta$.

\subsection{Quadratic Action For Scalar Perturbations}

\subsubsection{Flat FRW}

In the flat FRW case, the quadratic action for scalar perturbations is
\be\label{flatscalar}
S^{(2)}=-\int d^{4}x~ a^{2}\gamma~ \Big\{3{\zeta'}^{2}-6\mathcal{H}\zeta'A+(2A+\zeta)D^{2}\zeta+a^{2}V A^{2}
+2(\mathcal{H}A-\zeta')D^{2}B+2c\theta'\zeta D^{2}\lambda \Big\}~,
\ee
where $D^{2}=\gamma^{ij}D_{i}D_{j}$ and $\gamma=\det(\gamma^{a}_{~i})=\sqrt{\det(\gamma_{ij})}$.
One can see that none of $\lambda$, $B$ and $A$ is dynamical, they induce the following constraints:
\bea
& & \zeta=0~,\label{Cflat1}\\
& & \zeta'-\mathcal{H}A=0~,\label{Cflat2}\\
& & -3\mathcal{H}\zeta'+D^{2}\zeta+a^{2}V A+\mathcal{H}D^{2}B=0~.\label{Cflat3}
\eea
Compared with GR, there is one more constraint (\ref{Cflat1}) here.
Substituting these constraints back into the action (\ref{flatscalar}), we get $S^{(2)}=0$.
Hence there is no scalar dynamical degree of freedom at the linear perturbation level, this is the same with the result of Ref.\,\cite{PVtele1}.
This result does not depend on the gauge, because $\zeta$ is gauge invariant.

\subsubsection{Closed FRW}

In the closed FRW background, the quadratic action for scalar perturbations is
\bea
S^{(2)}=&-&\int d^{4}x~ a^{2}\gamma~\Big\{3{\zeta'}^{2}-6\mathcal{H}\zeta'A+(2A+\zeta)D^{2}\zeta+(a^{2}V-3\mathcal{K}^{2}) A^{2}
+3(\mathcal{K}^{2}-3c\mathcal{K}\theta')\zeta^{2}\nonumber\\
&+&6\mathcal{K}^{2}A\zeta
+2(\mathcal{H}A-\zeta')D^{2}B-\mathcal{K}^{2}B D^{2}B+c\theta'(2\zeta+\mathcal{K}\lambda)D^{2}\lambda-c\mathcal{K}\theta'\beta D^{2}\beta
\Big\}~.\label{closedscalar}
\eea
One can see that $\beta$, $\lambda$, $A$ and $B$ are all non-dynamical fields.
All these non-dynamical fields induce the following constraints:
\bea
& &\beta=0~,\label{Cclosed1}\\
& &\zeta+\mathcal{K}\lambda=0~,\label{Cclosed2}\\
& &\zeta'-\mathcal{H}A+\mathcal{K}^{2}B=0~,\label{Cclosed3}\\
& & -3\mathcal{H}\zeta'+D^{2}\zeta+3\mathcal{K}^{2}\zeta+(a^{2}V-3\mathcal{K}^{2})A+\mathcal{H}D^{2}B=0~.
\label{Cclosed4}~
\eea
Compared with GR, there are two more constraints (\ref{Cclosed1}) and (\ref{Cclosed2}) here.
These two constraints are used to solve $\beta$ and $\lambda$ respectively.
And the constraints (\ref{Cclosed3}) and (\ref{Cclosed4}) can be used to solve $A$ and $B$.
So it can be expected that there will be one scalar dynamical degree of freedom $\zeta$ in the end.
But it is worth noting that when $K=0$, that is, the constraint (\ref{Cclosed2}) will become $\zeta=0$. This returns to the case of $K=0$ in the previous subsection.

In closed FRW universe, the spatial line element can be expressed as $d\hat{s}^{2}=\gamma_{ij}dx^{i}dx^{j}=(1/K)[d\chi^{2}+\sin^{2}\chi(d\varphi^{2}+\sin^{2}\varphi\, d\phi^{2})]$.
Then, the hyperspherical harmonics $\mathcal{S}_{nlm}(\vec{x})$ (for more details, see \cite{spherescalar1,spherescalar2}) can be expressed as
\be
\mathcal{S}_{nlm}(\vec{x})=K^{\frac{3}{4}}\sin^{l}\chi\, \frac{d^{l+1}(\cos n\chi)}{d(\cos\chi)^{l+1}}\,Y_{lm}(\varphi, \phi)~,
\ee
where $Y_{lm}(\varphi, \phi)$ is the spherical harmonic, $n$ is a positive integer, $l$ is an integer satisfying $0\leq l\leq (n-1)$ and $m$ is an integer satisfying $|m|\leq l$.
These hyperspherical harmonics $\mathcal{S}_{nlm}(\vec{x})$ constitute a complete and orthogonal set for the expression of a scalar field on three-sphere.  These hyperspherical harmonics satisfy the following properties:
\bea
& &\quad\quad D^{2}\mathcal{S}_{nlm}(\vec{x})=-K(n^2-1)\mathcal{S}_{nlm}(\vec{x})~,\\
& &\int d^{3}x\, \gamma\, \mathcal{S}_{nlm}(\vec{x})\mathcal{S}^{*}_{n'l'm'}(\vec{x})=\delta_{n n'}\delta_{l l'}\delta_{m m'}~.
\eea

In order to solve the constraints and finally obtain the quadratic action after the constraints are released,
we use hyperspherical harmonics $\mathcal{S}_{nlm}(\vec{x})$  to expand the perturbation variables.
\be
\zeta(\eta,\vec{x})=\sum^{\infty}_{n\geq2}\sum^{n-1}_{l=0}\sum^{l}_{m=-l}\zeta_{nlm}(\eta)\mathcal{S}_{nlm}(\vec{x})~.
\ee
Note that the sum starts at $n=2$ since the $n=1$ mode is homogeneous and thus a part of the background.
Then apply the same expansions to the perturbation variables $A$ and $B$ which represent constraints, and substitute these constraints back into the action (\ref{closedscalar}), we get the final form of the quadratic action for scalar perturbations:
\be
S^{(2)}=\sum^{\infty}_{n\geq2}\sum^{n-1}_{l=0}\sum^{l}_{m=-l}\int d\eta~ z^{2}
\left(\frac{1}{2}|\zeta_{nlm}'|^{2}-\frac{1}{2}\omega^{2}|\zeta_{nlm}|^{2}\right)~,
\ee
\be\nonumber
\text{where}~ \left\{ \begin{array}{lll}
                   \displaystyle {z^{2}=\frac{a^{2}\theta'^{2}}{\mathcal{H}^{2}}\left(1-\frac{\theta'^{2}}{\mathcal{F}_{n}}\right)}\\
                    \\
                    \displaystyle{\omega^{2}=k^{2}+\frac{4\mathcal{K}^{2}}{3\mathcal{F}_{n}}\left[a^{2}V_{n}-(n^{2}-1)\theta'^{2}\right]
                      +\frac{2c\mathcal{K}}{\theta'}\left(\mathcal{F}_{n}-12\mathcal{K}^{2}+\frac{4\mathcal{K}^{2}\theta'^{2}}{\mathcal{F}_{n}}\right)}
               \end{array}\right.
\ee
In above expression, $k^{2}=(n^{2}-1)\mathcal{K}^{2}$, $V_{n}=(n^{2}-4)(V+3\mathcal{H}V_{\theta}/\theta')$
and $\mathcal{F}_{n}=2(n^{2}-4)\mathcal{H}^{2}+\theta'^{2}$.
This quadratic action shows clearly that there is one dynamical scalar degree of freedom. It comes from the dynamical scalar field $\theta(x)$. We introduced its dynamics at the beginning.
So the loss of scalar dynamical degree of freedom is only a special phenomenon in the case of flat FRW background.
Finally, $\mathcal{F}_{n}\geq\theta'^{2}$, so $z^{2}\geq0$, thereby avoids the ghost instability.

\subsubsection{Open FRW}

In the case of open FRW background, the quadratic action for scalar perturbations is
\bea
S^{(2)}=&-&\int d^{4}x~ a^{2}\gamma~\Big\{3{\zeta'}^{2}-6\mathcal{H}\zeta'A+(2A+\zeta)D^{2}\zeta+(a^{2}V+3\mathcal{K}^{2}) A^{2}
-3\mathcal{K}^{2}\zeta^{2}\nonumber\\
&-&6\mathcal{K}^{2}A\zeta
+2(\mathcal{H}A-\zeta')D^{2}B+\mathcal{K}^{2}B D^{2}B+2c\theta'(\zeta-\mathcal{K}\beta)D^{2}\lambda
\Big\}~.\label{openscalar}
\eea
This action (\ref{openscalar}) has a very different form compared with the one  (\ref{closedscalar}) for closed universe. This is because the corresponding background solutions (\ref{closed}) and (\ref{open}) have much different forms. It can be seen that $\beta$, $\lambda$, $A$ and $B$ are all non-dynamical fields.
All these non-dynamical fields induce the following constraints:
\bea
& &\lambda=0~,\label{Copen1}\\
& &\zeta-\mathcal{K}\beta=0~,\label{Copen2}\\
& &\zeta'-\mathcal{H}A-\mathcal{K}^{2}B=0~,\label{Copen3}\\
& & -3\mathcal{H}\zeta'+D^{2}\zeta-3\mathcal{K}^{2}\zeta+(a^{2}V+3\mathcal{K}^{2})A+\mathcal{H}D^{2}B=0~.
\label{Copen4}
\eea
Similarly, in comparison with GR, there are two more constraints (\ref{Copen1}) and (\ref{Copen2}) here.
These two constraints can be used to solve $\beta$ and $\lambda$ respectively.
And the constraints (\ref{Cclosed3}) and (\ref{Cclosed4}) can be used to solve $A$ and $B$.
Again, when $\mathcal{K}=0$, the constraint (\ref{Cclosed2}) will become $\zeta=0$ and this returns to the case of flat universe.

In the open FRW universe, the spatial line element can be expressed as
$d\hat{s}^{2}=\gamma_{ij}dx^{i}dx^{j}=(1/K)[d\chi^{2}+\sinh^{2}\chi(d\varphi^{2}+\sin^{2}\varphi\, d\phi^{2})]$.
Then, the $D^{2}$ eigenfunctions  $\mathcal{Z}_{nlm}(\vec{x})$ (for more details, see \cite{openscalar}) can be expressed as
\be
\nonumber\mathcal{Z}_{nlm}(\vec{x})=K^{\frac{3}{4}}\frac{\Gamma(in+l+1)}{\Gamma(in)}\left(\sinh\chi\right)^{-\frac{1}{2}}
P^{-l-\frac{1}{2}}_{in-\frac{1}{2}}(\cosh\chi)\,Y_{lm}(\varphi,\phi)~,
\ee
where $\Gamma(x)$ is the gamma function, $P^{\alpha}_{\beta}(x)$ is the associated Legendre function of the first kind and $n$ is a positive real number.
These functions $\mathcal{Z}_{nlm}(\vec{x})$ constitute a complete and orthogonal set for
the expression of a scalar field on three-dimensional hyperboloid. These functions $\mathcal{Z}_{nlm}(\vec{x})$ satisfy the following properties:
\bea
& &\quad\quad D^{2}\mathcal{Z}_{nlm}(\vec{x})=-|K|(n^{2}+1)\mathcal{Z}_{nlm}~,\\
& &\int d^{3}x\,\gamma\, \mathcal{Z}_{nlm}(\vec{x})\mathcal{Z}^{*}_{nlm}(\vec{x})=\delta(n-n')\delta_{ll'}\delta_{mm'}~.
\eea

In order to solve the constraints and finally obtain the quadratic action after the constraints are released,
we use the $D^{2}$ eigenfunctions $\mathcal{Z}_{n lm}(\vec{x})$  to expand the perturbation, i.e.,
\be
\zeta(\eta,\vec{x})= \sum_{l=0}^{\infty}\sum_{m=l}^{l} \int_{0}^{\infty} dn~ \zeta_{n lm}(\eta){\mathcal{Z}}_{n lm}(\vec{x})~,
\ee
and make similar expansions for the perturbation variables $A$ and $B$ which represent constraints.
Then substitute these constraints back into the action (\ref{openscalar}), we get the final form of the quadratic action for scalar perturbations:
\be
S^{(2)}=\sum^{\infty}_{l=0}\sum^{l}_{m=-l}\int_{0}^{\infty} dn\int d\eta~ z^{2}
\left(\frac{1}{2}|\zeta_{nlm}'|^{2}-\frac{1}{2}\omega^{2}|\zeta_{nlm}|^{2}\right)~
\ee
\be\nonumber
\text{where}~ \left\{ \begin{array}{lll}
                   \displaystyle {z^{2}=\frac{a^{2}\theta'^{2}}{\mathcal{H}^{2}}\left(1+\frac{\theta'^{2}}{\mathcal{F}_{n}}\right)}\\
                    \\
                    \displaystyle{\omega^{2}=k^{2}+\frac{4\mathcal{K}^{2}}{3\mathcal{F}_{n}}\left[(n^{2}+1)\theta'^{2}-a^{2}V_{n}\right]}
               \end{array}\right.
\ee
and $k^{2}=(n^{2}+1)\mathcal{K}^{2}$, $V_{n}=(n^{2}+4)(V+3\mathcal{H}V_{\theta}/\theta')$
and $\mathcal{F}_{n}=2[3\mathcal{K}^{2}+(n^{2}+1)\mathcal{H}^{2}+a^{2}V]$.
This quadratic action also shows clearly that there is one dynamical scalar degree of freedom.
Finally, for the case of $V(\theta)\geq0$ as in most inflationary models, we always have $\mathcal{F}_{n}>0$, and thus $z^{2}>0$, thereby the quadratic action avoid the ghost instability.

\subsection{Quadratic Actions For Tensor Perturbations}

\subsubsection{Flat FRW}
In the flat FRW case, we can use the circular polarization bases $\hat{e}^{L}_{ij}$, $\hat{e}^{R}_{ij}$ and plane wave $e^{i\vec{k}\cdot\vec{x}}$ to expand tensor perturbation
\begin{equation}
h_{i j}^{T}(t, \vec{x})=\sum_{A} \int \frac{d^{3} k}{(2 \pi)^{3 / 2}} h^{A}(t, \vec{k}) \hat{e}_{i j}^{A}(\vec{k}) e^{i k_{j} x^{j}}~,
\end{equation}
The bases satisfy the relation: $n^{l} \varepsilon_{li}^{~\,k} \hat{e}_{j k}^{A}=i \lambda_{A} \hat{e}_{i j}^{A}$,
here $A=L,R$ and $\lambda_{L}=-1$, $\lambda_{R}=1$, $\vec{n}$  is the unit vector of $\vec{k}$.
The final form of the quadratic action for tensor perturbation is
\be\label{flattensor}
S=\sum_{A} \int d \eta d^{3} k~ \frac{a^{2}}{4}\left[h^{A *^{\prime}} h^{A^{\prime}}-\left(k^{2}+c \theta^{\prime} \lambda_{A} k\right) h^{A *} h^{A}\right]~.
\ee
The quadratic action (\ref{flattensor}) is exactly the same as the one obtained in Ref.\,\cite{PVtele1}.
This is because the tensor perturbations only come from the tetrad, they have no origin from the spin connection or the local Lorentz matrices.
 This action shows clearly that there is no ghost instability, contrary to the CS gravity.
And it can be seen that the modified dispersion relation
$\omega_{A}^{2}=k^{2}+\lambda_{A} c \theta^{\prime} k=k^{2}\left(1+\lambda_{A} c \theta^{\prime} / k\right) \equiv k^{2}\left(1+\mu_{A}\right)$ is helicity dependent.
In terms of the notation of Ref.\,\cite{Zhao:2019xmm}, the deviation from the standard dispersion relation is characterized by
the parameter $\mu_{A}$.  Consider small coupling and slow evolution of $\theta$, one can find that GWs with different helicities will
have different phase velocities $v_{p}^{A}=\omega_{A} / k \simeq 1+\lambda_{A} c \theta^{\prime} /(2 k)$  and same group velocity
$v_{g}=d \omega_{A} / d k \simeq 1+c^{2} \theta^{\prime 2} /\left(8 k^{2}\right)$ up to the order $\mathcal{O}\left(c^{2}\right)$.
This is the so-called velocity birefringence phenomenon of GWs, which is a direct reflection of the parity violation in this model. The velocity birefringence here is very similar to the cosmic
birefringence induced by electromagnetic Chern-Simons coupling \cite{ECS1,ECS2,ECS3,ECS4}.
Since the deviation $\mu_{A}$ is inversely proportional to $k$, this is an infrared effect, contrary to most PV gravity models in the literature. We can also see that the
phase velocity difference or the deviation of the group velocity from the speed of light in vacuum become important
only at the region of small $k$ (large scales). Besides the velocity birefringence, the amplitudes for both helicities of GWs are the same, this is different from the
CS modified gravity either.
It is expectable that from the quadratic action (\ref{flattensor}), the produced primordial GWs in the early universe will
have different power spectra for left- and right-handed polarizations. This will result in the correlation between the
E- and B-modes polarizations of the cosmic microwave background radiation (CMB), and may have effects on the planned CMB experiments \cite{CMB1,CMB2}.

\subsubsection{Curved FRW}

In the curved FRW case, we can use the symmetric traceless and transverse tensor harmonics $\{Q^{(nlm)}_{ij}(\vec{x}), \bar{Q}^{(nlm)}_{ij}(\vec{x})\}$ (for more details, see \cite{Challinor:1999xz}) to expand the tensor perturbations. Below we briefly introduce how to construct the symmetric traceless and transverse tensor harmonics.

In the three-dimensional maximum symmetric space $d\hat{s}^{2}=\gamma_{ij}dx^{i}dx^{j}$, starting at some arbitrary point $O$,
we construct all geodesics with unit tangent vectors
$b^{i}(\chi)$, where $\chi\geq0$ (the equality holding at $O$) is the parameter of the geodesic.
A vector field $b^{i}$ and a scalar field $\chi$ can then be constructed.
Next, we consider the rank-$l$ symmetric trace-free tensor field $\mathcal{Q}_{i_{1}...i_{l}}$ satisfying $b^{j}D_{j}\mathcal{Q}_{i_{1}...i_{l}}=0$.
Since the $\mathcal{Q}_{i_{1}...i_{l}}$ have only $2l+1$ degrees of freedom, it is convenient to introduce a set of orthogonal
basis tensor $\{\mathcal{Q}^{(l m)}_{i_{1}...i_{l}}\}$, with $m=-l,...,l$.
For convenience, we define $\mathcal{Q}^{(lm)}\equiv\mathcal{Q}^{(l m)}_{i_{1}...i_{l}}b^{i_1}...b^{i_l}$,
$\mathcal{Q}^{(lm)}_{i}\equiv\mathcal{Q}^{(l m)}_{i,j_{2}...j_{l}}b^{j_2}...b^{j_l}$ and $\mathcal{Q}^{(lm)}_{ij}\equiv\mathcal{Q}^{(lm)}_{i,j,k_{3}...k_{l}}b^{k_3}...b^{k_l}$.
We also define $\mathcal{H}_{ij}\equiv\gamma_{ij}-b_{i}b_{i}$ and
$[A_{ij}]^{TT}\equiv (\mathcal{H}^{k_1}_{i}\mathcal{H}^{k_2}_{j}-\mathcal{H}_{ij}H^{k_{1}k_{2}}/2)A_{k_{1}k_{2}}$.
Then the symmetric traceless and transverse tensor harmonics can be expressed as
\bea
& & \displaystyle{Q^{(nlm)}_{ij}(\vec{x})=T^{(nl)}_{1}(\sqrt{|K|}\chi)\left(b_{i}b_{j}-\frac{1}{2}\mathcal{H}_{ij}\right)\mathcal{Q}^{(l m)}
-T^{(nl)}_{2}(\sqrt{|K|}\chi)b_{(i}\mathcal{H}^{k_{l}}_{j)}\mathcal{Q}^{(l m)}_{k_l}+T^{(nl)}_{3}(\sqrt{|K|}\chi)[\mathcal{Q}^{(l m)}_{ij}]^{TT},
}\\
& &\bar{Q}^{(nlm)}_{ij}(\vec{x})=\bar{T}^{(nl)}_{1}(\sqrt{|K|}\chi)[b_{k_1}\varepsilon^{k_{1}k_{2}}_{~~~~(i}\mathcal{Q}^{(l m)}_{j)}]^{TT}
+\bar{T}^{(nl)}_{2}(\sqrt{|K|}\chi)b_{k_1}b_{(i}\varepsilon_{j)}^{\ k_{1}k_{2}}\mathcal{Q}^{(l m)}_{k_{2}},
\eea
where
\bea
& & \nonumber\displaystyle{ T^{(nl)}_{1}(x)= \frac{\Phi^{n}_{l}(x)}{S_{K}^{2}(x)}~,
}\\
& & \nonumber\displaystyle{ T^{(nl)}_{2}(x)=\frac{-2}{(l+1)S_{K}^{2}(x)}\frac{d}{dx}[S_{K}^{3}(x) T^{(nl)}_{1}(x)]~,
}\\
& & \nonumber\displaystyle{ T^{(nl)}_{3}(x)=\frac{-l}{l+2}T^{(nl)}_{1}(x)-\frac{1}{(l+2)S_{K}^{2}(x)}\frac{d}{dx}[S_{K}^{3}(x) T^{(nl)}_{2}(x)]~,
}\\
& & \nonumber\displaystyle{ \bar{T}^{(nl)}_{2}(x)=\frac{-2n}{l+1}\frac{\Phi^{n}_{l}(x)}{S_{K}(x)}~,
}\\
& &  \nonumber\displaystyle{ \bar{T}^{(nl)}_{1}(x)= \frac{-1}{(l+2)S_{K}^{2}(x)}\frac{d}{dx}[S_{K}^{3}(x) \bar{T}^{(nl)}_{2}(x)]~,
}
\eea
and $\Phi^{n}_{l}(x)$ is the ultra-spherical Bessel function.
When $K>0$, $S_{K}(x)=\sin(x)$, $n\geq3$ is an integer, $l$ is an integer satisfying $2\leq l\leq (n-1)$ and $m$ is an integer satisfying $|m|\leq l$.
When $K<0$, $S_{K}(x)=\sinh(x)$, $n$ is a positive real number, $l\geq2$ is an integer and $m$ is an integer satisfying $|m|\leq l$.

For convenience, we recombine these  symmetric traceless and transverse tensor harmonics into the circular polarization bases:
\be
(Q^{A}_{nlm})_{ij}=\mathcal{N}_{nl}\left[Q^{(nlm)}_{ij}-\lambda_{A}\bar{Q}^{(nlm)}_{ij}\right]~,
\ee
 where $\mathcal{N}_{nl}$ is a  normalization coefficient, $A=L,R$ and $\lambda_{L}=-1$, $\lambda_{R}=1$.
When $K>0$, we define $k=\sqrt{\mathcal{K}(n^{2}-3)}$; and for $K<0$,  $k=\sqrt{\mathcal{K}(n^{2}+3)}$.
It can be proved that these bases satisfy the following relations:
\be
D^{2}(Q^{A}_{nlm})_{ij}=-k^{2}(Q^{A}_{nlm})_{ij}~,~~
\varepsilon_{kl(i}D^{k}(Q^{A}_{nlm})^{l}_{~j)}=-\lambda_{A}k\sqrt{1+\frac{3K}{k^{2}}}\,(Q^{A}_{nlm})_{ij}~.
\ee
We can always choose an appropriate normalization coefficient $\mathcal{N}_{nl}$ so that the bases are normalized to
\be
\int d^{3}x~\gamma~ (Q^{A}_{nlm})_{ij}(Q^{B}_{n'l'm'})^{ij}=\left\{
           \begin{array}{lll}
           \displaystyle{2\delta^{AB}\delta_{nn'}\delta_{ll'}\delta_{mm'}} & ~(K>0)\\
           \\
           \displaystyle{2\delta^{AB}\delta(n-n')\delta_{ll'}\delta_{mm'}} & ~(K<0)
           \end{array}\right.
\ee
The tensor perturbations can be expanded as
\be
h^{T}_{ij}(\eta,\vec{x})=\sum_{A=L,R}\sum_{{k}}h^{A}_{nlm}(\eta)(Q^{A}_{nlm})_{ij}(\vec{x})~,\quad\text{where}~
\sum_{{k}}\equiv\left\{
           \begin{array}{lll}
           \displaystyle{\sum_{n=3}^{\infty}\sum_{l=2}^{n-1}\sum_{m=-l}^{l}}  & ~(K>0)\\
           \\
           \displaystyle{\sum_{l=2}^{\infty}\sum_{m=-l}^{l} }\int^{\infty}_{0}dn & ~(K<0)
           \end{array}\right.
\ee
Hence, we get the final form of the quadratic action for tensor perturbation
\be\label{curvedtensor}
S^{(2)}=\sum_{A=L,R}\sum_{k}\int d\eta~\frac{a^{2}}{4}
        \left[(h^{A^{\prime}}_{nlm})^{2}-\omega_{A}^{2}(h^{A}_{nlm})^{2}\right]~,
\ee
where
\be\nonumber
\omega_{A}^{2}=
        \left\{
        \begin{array}{lll}
           \displaystyle{k^{2}+2\mathcal{K}^{2}+c\theta'\mathcal{K}+\lambda_{A}c\theta'k\sqrt{1+\frac{3\mathcal{K}^{2}}{k^{2}}}} & ~(K>0)\\
           \\
           \displaystyle{k^{2}-2\mathcal{K}^{2}+\lambda_{A}c\theta'k\sqrt{1-\frac{3\mathcal{K}^{2}}{k^{2}}}} & ~(K<0)
           \end{array}\right.~
\ee
The quadratic action (\ref{curvedtensor}) shows no ghost instability.
Again, it produces the velocity birefringence phenomenon rather than the amplitude birefringence phenomenon.
If the three curvature is very small, $\mathcal{K}\approx0$, the space of the background is approximately flat,
the modified dispersion relation is approximately $\omega_{A}^{2}\approx k^{2}+\lambda_{A} c \theta^{\prime} k$, returning to the dispersion relation obtained in the universe with spatially flat FRW background.

\section{conclusion}\label{conclusion}

In this paper we revisited the recently proposed parity violating gravity model \cite{PVtele1}, which is based on the teleparallel gravity theory and has a simple form. More importantly it is ghost free.  In this revisit, we gave up the Weitzenb\"{o}ck connection and restored the local Lorentz symmetry of this model. With the full local Lorentz covariance, this model is not a pure tetrad theory any more. We applied this new version of the model to the universe which has a general FRW background. We studied the background evolution and the linear cosmological perturbations. We found that in the case of flat FRW background the background evolution is the same as GR, but the scalar field $\theta(x)$ which couples to gravity through the PV term is not counted as a dynamical degree of freedom at the linear perturbation level, and the tensor perturbations present velocity birefringence in stead of amplitude birefringence phenomenon. These confirmed the results obtained in  \cite{PVtele1}. We also considered the cases of curved FRW backgrounds. We found that, the PV modification only affects the background evolution of the closed universe, in both the open and closed universes the fluctuation of the scalar field $\theta$ remained as a dynamical degree of freedom, the tensor perturbations are similar to those in flat universe except that in the curved universe their dispersion relations have further dependences on the spatial curvature.
By the way, the phenomenon of degrees of freedom being hidden under special backgrounds also appears in $f(\mathbb{T})$ models \cite{Ong:2013qja} and the models of massive gravity \cite{DeFelice:2012mx}.

{\it Acknowledgement}: This work is supported by NSFC under Grant No. 12075231, 11653002, 12047502 and 11947301.


{}


\end{document}